# Searching for short baseline anomalies with the LAr-TPC detector at shallow depths.

Carlo Rubbia

GSSI-INFN, GranSasso Science Institute, L'Aquila, Italy

Abstract.

The ICARUS Collaboration has operated successfully the Liquid Argon time projection chamber (LAr-TPC) [6], a novel and continuously sensitive "bubble chamber like" neutrino detector in the GranSasso Laboratory and an underground neutrino beam coming from the CERN-SPS.

ICARUS may now be moved at the 8 GeV FNAL-Booster for a search of LSND-like neutrino-electron anomalies at a shallow depth and shorter distance from the target, where three experiments will simultaneously study neutrinos at three different locations. New and substantial modifications are described in order to make ICARUS operable in the presence of such a large cosmic ray muon background.

*1. Introduction.*

Neutrino oscillations have established a picture consistent with the mixing of three physical neutrino ν–e, ν–μ and ν–τ with mass eigen-states $\nu_1$, $\nu_2$ and $\nu_3$. Mass values are so far unknown. Mass differences turn out to be relatively small, with $\Delta m^2_{31} \approx 2.4 \times 10^{-3}$ eV$^2$ and $\Delta m^2_{21} \approx 8 \times 10^{-5}$ eV$^2$. Over the last several years, neutrinos have been the origin of an impressive number of "surprises" and more might be ahead. For instance: neutrinos are Dirac fermions like quarks or maybe Majorana particles instead, in which the antineutrino coincides with the right handed neutrino and with a neutrino-less lepton $\Delta L = 2$ violation; is there more physics beyond the standard model; are there "anomalies" that, provided confirmed experimentally, might be due to the presence of additional, larger squared mass differences; are there some additional "sterile" neutrinos; or maybe even other totally unexpected effects. The astronomical importance of neutrinos from Space is immense, so is their role in the cosmic evolution. A more massive addition to ordinary neutrinos will necessarily influence the Dark Matter.

Sterile neutrinos [1] are hypothetical types of neutrinos that do not interact via any of the fundamental interactions of the Standard Model, except gravity. Since they would not interact electromagnetically, weakly, or strongly, they are extremely difficult to detect. However, oscillations with ordinary neutrinos are possible in order to reveal their presence.

Several different neutrino related "anomalies" have been experimentally reported, namely (1) an apparent disappearance signal in the antineutrino spectra detected from near-by nuclear reactors [2], (2) the calibrations with Mega-Curie k-capture sources [3] in experiments detecting solar ν–e and (3) the observation of presumed excess signals of electrons in neutrinos and in antineutrino experiments from particle accelerators [4].

A neutrino disappearance signal [2] has been observed at a few meters away from the nuclear reactors. It is characterized by a ratio R between the detected and the expected rates slightly smaller than 1, namely R = 0.927 ± 0.023. However, different reactors experiments rely on the same flux prediction. Recent evaluations have however increased the expected flux to R = 0.935 → 0.958 [± 0.009(stat) ± 0.027(syst)], with a surviving deviation from R =1 of 2.7 to 1.4 standard deviations, depending on the model for the flux predictions.

The calibration signal produced by intense artificial k-capture sources of $^{51}Cr$ and $^{37}Ar$ have been recorded by the SAGE and GALLEX experiments [3]. The combined ratio between the detected and the neutrino rates predicted by the source is R = (0.86 ± 0.05), about 2.7 standard deviations from R = 1.

The LSND experiment at Los Alamos National Laboratory [4], designed to search with high sensitivity for oscillations from $\mu^+$ decay at rest with a 1 mA proton intensity and 800 MeV energy, has reported an apparent excess signal of electrons from anti-neutrinos (and later also from neutrinos from $C^{12}$). Almost all events arise from positive decays since $\pi^-$ and $\mu^-$ are readily captured in the beam stop. The production of k-decays is negligible. The signature is $\bar{v}_e + p \rightarrow e^+ + n$, followed by $n + p \rightarrow d + Y\ (2.2\ MeV)$, where Y is a 2.2 MeV photon correlated with a positron. The result is an excess of ( 87.9 ± 22.4 ± 6 ) events, which if due to neutrino oscillations corresponds to their local oscillation probability of $(0.264 ± 0.067 ± 0.045)\%$. The v-e rate is calculated to be only $4 \times 10^{-4}$ relative to v-μ in the 36 < $E_v$ < 52.8 MeV energy interval. The observation of a significant v-e rate is therefore considered as a 3.8 σ evidence for $v\mu \rightarrow ve$ oscillations.

A further search for an excess of electron neutrino events for $E_v/L \approx 1\ (MeV/m)$ in the v and $\bar{v}$ beams at the FNAL 8 GeV proton Booster and $\langle E_v \rangle \approx 800\ MeV$ has been performed by the MiniBooNe experiment [5]. The LSND anomaly has not been fully confirmed, but a *new anomaly* is observed at $E_v < 470\ MeV$ both for v and $\bar{v}$ events, to be further clarified in a number of future experiments.

The ICARUS experiment CNGS-2 with the help of a LAr-TPC [6] has strongly reduced the surviving window of opportunity for the LSND anomaly [7]. All three independent signals may now be due to the possible existence of at least one fourth non standard and heavier neutrino state, driving oscillations at a small distance, of the order of $\Delta m^2_{new} \approx 1\ eV^2$ and a relatively small mixing angle $\sin^2(2\theta_{new})$.

In these previous short baseline experiments the electron detection has been performed with the help of scintillation and Cerenkov light collected at the boundaries of the detector volume. The signal of the events can be considerably improved with the introduction of the Liquid Argon Time Projection Chamber (LAr-TPC), the new "imaging" and continuously sensitive detector developed by the ICARUS Collaboration [6], capable to identify unambiguously all reactions with a resolution of few $mm^3$, comparable to the one of a classic bubble chamber. The collaboration has already successfully operated over the last three years the T600 detector with a LAr mass of 760 t underground in the Hall B of the GranSasso laboratory and neutrinos coming from CERN. The present paper describes a new experiment in which the LAr-TPC's will be located near the surface and at a relatively short distance from a neutrino target.

With the observation of the neutrino oscillations at several suitable $E_v/L$ distances from the target such as to ensure optimal matching, the novel LAr-TPC technology introduces important new features, which should allow a definitive clarification of all the above described "anomalies". In this way the values of the relevant mass difference $\Delta m^2$ and of the mixing angle $\sin^2(2\theta)$ for the additional neutrino oscillation can be identified for both $v$ and $\bar{v}$ focussed beams. Both ν-e and ν–μ components are cleanly identified. Two (or more) LAr-TPC's of substantial mass, both at far and near distance from the FNAL Booster will provide the very high rates in order to record relevant distance dependent effects at the few % level ($>10^6$ ν–μ, $>10^3$ ν-e).

A magnetic field of the order of 1 Tesla should be added as a second phase to the LAr-TPC in order to extend the programme to a pure anti-neutrino beam. The opportunity of determining the final charge of the produced lepton is justified by the very large contamination of neutrino events in the anti-neutrino beam, at the Booster about three times more abundant than the anti-neutrinos events.

It is presently planned to reconfigure ICARUS and to increase the data taking rate by doubling its volume, going from T600 to T1200 and therefore approaching to a sensitive mass of about 1000 ton. This is made possible with a relatively small effort maintaining essentially the existing number of active elements and simply enlarging the electron drift distance of the individual readout gaps from the original 1.5 m to slightly more than 3 m. This is now possible since the free electron lifetime of the high purity LAr has been brought by us to an exceptional value of 18 m [8].

A new experiment to search for ν-e anomalies is presently under joint consideration at FNAL by three different teams and with three distinct but rather similar LAr-TPC detectors. It will be composed of the ICARUS-T1200, Microboone and LAr1—ND detectors at distances of 600 m, 470 m and 110 m from the BNB target, total masses of 1600 t, 170 t and 180 t and LAr active masses of 1000 t, 89 t and 82 t respectively. Here we shall concentrate only on the required modifications for the ICARUS detector. However, it is very likely that similar considerations may necessarily have to be applied also to the other detectors.

2. *Implications of the operation of a large LAr-TPC at shallow depths.*

As described in [6], the LAr-TPC is capable of collecting with high resolution all ionizing tracks within the sensitive volume located between a very wide (up to several meters) cathode at high voltage and the anode structure made of several closely located wire planes near ground and oriented at different angles, orthogonal to the electric field. The lifetime of free electrons in the high purity LAr of ICARUS has recently attained 12 ms, corresponding to a free electron path of as much as 18 meters for a nominal field of 500 V/cm [8]. Two readout wire planes with a typical wire pitch of 3 mm record inductively the passing electron signals, followed by a final charge collecting wire plane. The orientations of the three planes are +60°, -60° and 0°. As well known, in any TPC each of the wire planes records only the time sequence of their *final arrivals collected at the end of the long drift path*. The full image of the event occurring after a "trigger" signal is then only subsequently reconstructed in 2D in each of the views combining the drift time distributions with the individual readout wires.

A renovated photo-multiplier arrangement located at the edges of the sensitive volume will collect the UV scintillation signal from the free electrons which is present in the LAr simultaneously to the ionization, wave-shifting the transparent liquid from the UV (about 128 nm) to the visible light with ordinary photo-multipliers. The scintillation light from LAr has two components, one of few ns, which can be used as a prompt signal and a slower one of ≈1.6 $\mu$s average duration. The photo-multipliers may also collect a significant, though much smaller amount of Cerenkov light. This remains as the dominant signal in the absence of UV conversion.

The detection process in the LAr-TPC is initiated by the "trigger" signal, as the prompt response to the occurrence of an "event". The "trigger" signal generated by the scintillation of the photo-multipliers is initiating the opening of a long "imaging" readout window, in which tracks will be recorded in a time sequence, collected serially by the readout planes while the electrons travel towards the end of the long drift path. The full image of the event is therefore progressively extracted from the drift time distributions and from the many readout wires.

The ICARUS detector has already successfully operated in the low background, deep underground conditions of the Hall B of the GranSasso Laboratory. The "trigger" was generated combining the light of each of the two T300 volumes as a coincidence behind the two opposite readout anodes at 3 m distance. Such a double coincidence has been necessary in order to select the signal above a prescribed level and in particular to remove the local β radioactivity of $^{39}$Ar, of the order of $10^6$ *c/s* for the full LAr volume of 760 ton. In the underground conditions of the CNGS experiment [7] this has been possible, since a single prompt "trigger" has always ensured the unique timing connection to the main "image" of the event.

However, the situation will be substantially different for a detector of this magnitude if placed at a shallow depth (a few metres deep), since several additional and uncorrelated cosmic ray related scintillation "triggers" will then be generally occurring continuously and at different times during the duration of one common "imaging" readout window. This represents a new problem since in order to reconstruct the true position of the track it is necessary to associate precisely the different timings of each the elements of the image to their own specific delay in the trigger.

The ICARUS measurements on surface during 2001 have shown that in the T600 with 476 t of visible LAr and during the 1 ms long "imaging" window, there are on the average as many as 24 muon tracks, in good agreement with the expected cosmic ray flux. Several of them were multiple muon "bundles" with several parallel tracks. Since the scintillation light from the photo-multipliers is also recorded from the regions without "imaging", the total number of muon related "trigger" signals, roughly extending their number like the volume, is expected to be of the order of 760 t/476 t x 24 = 38 trigger/ms, or on average a cosmic ray generated scintillation "trigger" every 26.3 μs.

The specific search of the LSND like anomalies at a shallow depth is based on the search of a signal with the presence of a singly ionizing neutrino induced electron or positron. High energy cosmic muons creating secondary showers may also produce singly ionizing background electrons or positron with similar energies. The distance between the locations of the muon inducing the shower and of the related γ –ray conversion is often of the order of many tens of centimetres (the radiation length in LAr is 14 cm) and therefore such a correlation cannot be automatically associated. Therefore, while the case of a beam associated neutrino event is well localized and hence correctly located by the unique beam trigger, the accidental background production of a cosmic ray generated γ–ray conversion is generally arbitrarily located in time with respect to the beam related initiating trigger. Due to the inevitable presence of many "triggers" the situation is very complex and not easily and unambiguously disentangled. In principle any of them could have generated a given observed positive or negative electron. A very sophisticated procedure is necessary in order to recognize beam related from accidental random events.

3. *Expected ICARUS operation at the BNB Booster.*

The neutrino facility in association with the 8 GeV FNAL-Booster provides a high purity ν-μ beam peaked around ~0.5 GeV, with an intrinsic ν-e contamination at ~0.5 % level. Usually ~$4 \times 10^{12}$ protons are delivered to the beryllium target in 1.6 μs long spills, corresponding for each nominal year of data taking to about $2.2 \times 10^{20}$ p.o.t. (protons on target), to about $5 \times 10^7$ beam pulses/y (y=year) and an over-all beam on live time of about 90 s/y. The main goal of the experiment is to identify conclusively the presence or absence of a LSND neutrino-electron signal at least at a 5 sigma level in a running period of about three years.

After the foreseen improvement programme to be carried out at CERN, it is assumed that the ICARUS detector should double its mass with ≈1.6 kton of very high purity LAr and ≈1 kton of fiducial mass. New Aluminium containment vessels will have to be constructed. The original four wire planes located near a grounded anode and their associated readout structure will be left essentially untouched, simply enlarging to slightly more than 3 m the distance of the anode to the cathode, which will have now to operate at twice the voltage, i.e. 150 kV, already successfully tested by the ICARUS team with the T600.

In order to be transported conveniently, the new 1 kton ICARUS detector will be made out of four rather than two separate new vessel units, each one of roughly the same external dimensions as the ones of the present T300 units. These four detectors will be stacked as two sets of two identical T600 like units, superimposed vertically above the pit (covered by 3 m of rock) to minimize the lateral dimensions of the building. Each of the four new T300 units will be followed by a ≈ 30 cm un-instrumented gap to bring the HV back to ground safely. Therefore the number of readout wires and of electronics channels will be preserved, however subdivided amongst four rather than two vessel units.

At the neutrino energies of the FNAL-Booster 60% of the events are quasi-elastic scatterings, the rest being single pion and deep-inelastic. At 600 m from the target and with a LAr active mass of 1 kton there are $3.3 \times 10^5$ $v_\mu CC/y$, namely about one event each 150 beam triggers. The calculated NC rate is $1.1 \times 10^5$ $v_\mu NC/y$.

The intrinsic v-e contamination occurs at the very low rate of 1780 $v_e/y$ (one event every 28100 bursts) for E ≤ 2 GeV and a possible LSND signal with 0.15% of the $v\mu CC$ namely $0.33 \times 10^6 \times 1.5 \times 10^{-3} = 495$ $v_e/y$. In the optimal energy interval 0.35 to 1.5 GeV the fractions are 0.81 for the $v\mu CC$ is and 0.72 for the $veCC$.

On the other hand, the cosmic ray background is very prolific of events. At the surface, the original T600 (two T300 units side to side) recorded a cosmic ray scintillation "trigger" every 26.3 µs. In a pit covered by 3 m of rock, muon rates are expected to be roughly one half of these values, namely one trigger every 52.6 µs or a trigger rate of 19 kc/s. Since there are now two detectors units that are superimposed vertically, there will be a partial overlap of the units and a substantial fraction of the muons will cross more units. As a result we anticipate under 3 m rock cover that a fast "trigger" scintillation signal due to cosmic rays hitting at least one of the four T300 units should occur every about 35 µs namely a trigger rate of 28.5 kc/s.

This fast "trigger" scintillation ([B] signal trigger) observed by the photo-multipliers from the LAr, must be put in coincidence with the 1.6 µs beam trigger ([A] signal) occurring at the rate of $5 \times 10^7$ beam pulses/y. A fast coincidence [AB] will occur at every 35 µs/1.6 µs = 21.9 beam pulses, namely at the huge rate of $5 \times 10^7/21.9 = 2.28 \times 10^6$ [AB] c/y. The photo-multiplier signals from each of the four T300 units are collected separately and it is therefore possible to identify simply the [B} unit(s) where there has been the fast [AB] coincidence.

Several additional and uncorrelated cosmic rays driven scintillation "triggers" ([B] signals) will be generally occurring continuously at different times during the 2 ms long duration of each "imaging" readout window. In each one of them there will be on average as many as 2 ms/35 µs = 57 scintillation LAr signals spread over the "imaging" window of the future ICARUS 1 kt volume[1]. Of course only the specific T300 unit or units where there has been the fast [AB] coincidence should be considered.

---

[1] The comic ray fluxes used here are the ones relative to the position in Pavia, Italy. FNAL fluxes are slightly lower, since nearer to the magnetic pole.

It is concluded that at least in its original configuration the ICARUS LAr-TPC detector cannot perform a practical search for LSND-like anomalies at shallow depths, since the background signals of the order of 2.28 x $10^6$ [AB] c/y, coming from cosmic rays in coincidence with the 1.6 µs beam trigger ([A] signal) are much too frequent.

Very effective new methods must be introduced in order to reduce the cosmic ray related signals and eventually on the same time also the additional beam related neutrino events produced outside and entering in the outer boundaries of the detector.

To this effect we remark that the ICARUS LAr volume is presently not externally shielded in contrast with other similar detectors like S-K, Mini-BooNE and many others, where an effective, prompt anti-coincidence is surrounding the whole active volume.

In ICARUS, each of the four LAr filled large volumes is completely contained inside a wide metallic Aluminium cubic box at a ground potential. To overcome this situation, we can insert inside such a box, a few centimetres away from the edges, a large number of finely segmented (f.i. 20 x 20 $cm^2$, rectangular or otherwise) thin counter plates parallel to the Aluminium walls, in order to detect the presence of the dE/dx signals generated by the electrons in the LAr with a relatively modest electric field (f.i. ≈ 1 kV/cm) added between the readout plates and the Aluminium grounded walls. These are fast counters, not TPC's. In this way, the presence, approximate position, pulse size and prompt timing of all charged tracks from the counter plates will be recorded by their electric signals. The realisation of these segmented anti-coincidence counters on each of the four T300 volumes is relatively trivial on the bottom and on the four lateral sides. The top face is more delicate and it may require some additional considerations on the best way of achieving a realistic coverage.

Such 4π covering detectors ([C] signals) may record each charged particle traversing the outer boundaries of the LAr containers with an average ≥ 95% estimated coverage efficiency (i.e. 5% anti-coincidence missing probability) at each crossing. For instance, the tracks traversing both doubly stacked T300 units may be detected up to four times, entering and exiting the volumes and therefore provide a much higher anti-coincidence efficiency. Under a depth of 3 m of rock and out of the 2.28 x $10^6$ [AB] c/y, the fraction of cosmic ray muons which are either *coming to rest or decay inside a single T300 volume* is estimated to be the order of 15 %, corresponding to 3.42 x $10^5$ [AB] c/y. These events dominate the resulting surviving [$AB\bar{C}$] rate, since in the case of multiple traversals the survival probability is considerably smaller and it is neglected at this stage.

The surviving cosmic muon rate, with the help of these additional 4π counters is reduced to 2.28 x $10^6$ x 0.15 x 0.05 = 1.71 x $10^4$ c/y, for instance only about 3% of 4.4 x $10^5$ c/y, the dominant ν–µ beam neutrino rate inside the full 1 kton fiducial volume. Therefore such a proposed and relatively simple addition of a 4π anti-coincidence is very effective in bringing under control the number of cosmic ray related muons.

We remark that a substantial fraction of the beam related neutrino events may be also automatically removed by the $[AB\bar{C}]$ coincidence, since some of the generated secondary particles, and in particular the long beam muon secondaries may also cross the anti-coincidence boundaries of the $4\pi$ ($[C]$ signal).

The cosmic ray events should be preferentially oriented vertically while the beam related neutrino events are oriented horizontally away from the beam direction. However, at the relatively low neutrino energies of the FNAL-Booster the directionality factors of a real or an apparent singly ionizing positive or negative electron shower are substantially smeared out. These effects are neglected at this level.

The analysis of the neutrino energy of the $\nu_\mu CC$ events requires the knowledge of the momentum of the muon tracks, many of which may range out beyond the sensitive volume of the detector. The momentum of these muons has been determined in ICARUS with the help of the multiple scattering of the tracks [8]. This requires the very precise location of the tracks along the volume. At shallow depths, the presence of a relatively large number of very slow ion charges, moving at a few m/s speed (whose effects are proportional to the square of the anode to cathode distance) may introduce significant space charge related distortions of the electric field that must be overcome introducing additional layers of reasonably spaced electric shaping wires inside the volume, in order to stabilize the electric field. An additional wire plane should also be added in front of the first anode readout plane since otherwise the first wire induction plane would be left opened to the electric fields of the about 3 m wide drift gap.

4. *Searching for a neutrino associated ν-e signal.*

As already pointed out, inside the enlarged ICARUS with 1 kton of sensitive LAr, the intrinsic neutrino associated ν-e contamination is expected to occur only at the low rate of 1780 $\nu_e/y$ even before any detection inefficiency. A hypothetical, additional LSND signal may be expected with additional $\approx 495\ \nu_e/y$. Instead, the minimal number of neutrino associated events are those which are produced inside the visible volume, dominated with 3.3 x $10^5$ $\nu_\mu CC/y$, and 1.1 x $10^5$ $\nu_\mu NC/y$, in total about 5 x $10^5$ $ev/y$. The dominant type of the neutrino events will be quasi-elastic (60%), followed by single pion production (30%) and deep inelastic scattering (DIS) at about 10%.

These ν-e events coming from neutrinos may be confused with the prolific source of γ-ray conversions from cascades from high energy cosmic ray muons. The surviving cosmic ray related $[AB\bar{C}]$ "trigger" rate can however be brought down to a reasonable rate with the insertion of the additional $4\pi$ anticoincidence [C] counters and under 3 m thick rock shielding.

However, over the 2 ms "imaging" windows and the 1 kt volume there will be a random background of converted γ-rays, simulating beam associated ν-e events. Such large number of random generated γ-rays conversions is evidently always present *in any "imaging picture", either beam related or otherwise.*

The probability that a given high energy muon generates an induced γ-ray > 200 MeV inside the LAr is estimated to be of the order of 0.02. Over a 2 ms "imaging" window there are on average = 58 additional [B]-type randomly distributed scintillation "triggers" roughly subdivided amongst the four T300 volumes. However, since the dimensions of the ν-e signal are relatively small, generally only one of the four separately light collecting T300 volumes has to be retained. Any of the 58/4 = 14.5 muon driven triggers signalling the presence of LAr activity can contribute to the induced γ-ray events and therefore the order of 0.02 x 14.5 = 0.29 γ-rays conversions are expected on the average within the chosen T300 volume. The visible volume probably represents about ≈2/3 of the total and consequently about 0.29 x 2/3 = 0.19 γ-ray conversions are expected to be visible in each of the observed imaging pictures.

The probability of a Compton effect in which the γ is converted into an electron in LAr is of the order of 3%. Therefore the Compton electrons > 200 MeV, which could mimic the neutrino associated ν-e in the selected T300 volume are of one event every 1/(0.03 x 0.19) = 175 imaging pictures. However in the Compton effect the nuclear recoil will be absent. The added requirement of the presence of an unambiguous nuclear recoil star will therefore remove this background very efficiently, no doubt of at least a factor 100, leading to less than one event every 17'500 imaging pictures but with a prize of a somewhat reduced ν-e detection efficiency (about 70%).

Ordinary pair production coming from showering of high energy cosmic ray muons will produce in most of the cases easily distinguishable doubly ionized e.m. events. However, because of the relatively low energy of the BNB, not all events may clearly appear as neatly showering, double ionizing $e^+$-$e^-$ pairs. In few % of the events the $e^+$-$e^-$ pair will be strongly asymmetric, with the low energy leg well below an acceptable energy and leaving an apparent singly ionizing electron as the high energy leg.

Low energy electron events in LAr have been already considered in the study of solar ν–e underground. A 10 MeV electron in a LAr-TPC is still observable although with very large multiple scatterings, while a 5 MeV electron is below practical detectability. The nominal path length in LAr of an electron of 5 MeV is about 1.8 cm, corresponding at most to 6 wires. But the effective number of wires that are actually hit, in view of the irregularities in the track, is significantly shorter than the actual true electron path length. Two or three wires are not uncommon.

We presume that about 1% of the pairs have the lower leg compatible with either a positive or negative electron of ≤ 5 MeV. Events that could mimic the neutrino associated ν-e in the selected T300 volume are therefore of one event every 1/(0.01 x 0.19) = 526 imaging pictures. These very low energy electrons or positrons signalled with only a few wires have to be distinguished from a tiny nuclear recoil "star" with one or maybe more few MeV heavy tracks at the production point or maybe a short proton recoil coming from a quasi-elastic ν-e. For instance a proton recoil of 200 MeV/c has a kinetic energy of 21 MeV and a LAr track of about 6 mm, i.e. only about two readout wires. A 3 MeV typical nuclear evaporation proton recoil has a momentum of 75 MeV/c and it remains generally contained inside a single wire signal.

This is a main background which is much harder to control, although some further topological separation between e.m. and nuclear "stars" may be possible. A very soft electron will appear as a very tiny recoil, but being at a very low momentum, it will heavily scatter before coming to rest. The proton recoil(s) will instead be very straight. Proton recoils are characterized by large angles and often even in the direction opposite with respect to the e.m. track. It is assumed that such a safe separation may be possible in some instances, however it needs to be more precisely studied with actual events.

In conclusion, the very asymmetric $e^+$-$e^-$ pairs represent a background that must be further reduced in order to be able to exploit the genuine beam associated ν-e signal. We remark that already for the minimal number of 5 x $10^5$ *ev/y* beam associated events, the indicated rate of one fake event every 526 imaging pictures would introduce about 950 *ev/y* in the selected T300 volume. In reality the actual number of neutrino related events will be inevitably a few times larger and so will be the background. An indicative background value of about 2'000 to 3'000 *ev/y* is more realistic at this stage.

Therefore further methods have to be added in order to reduce the number of imaging pictures, still maintaining a major fraction of the ν–e associated events:

1) The probability of having in the same image more than one candidate event is small enough to reject the event. Therefore we should exclude from the analysed sample the $\nu_\mu$–CC and the $\nu_\mu$–NC events, even if only partially visible inside the T1200 volume. While the CC events are easy to detect and to remove, the presence of NC is more delicate. The cosmic ray related 1.71 x $10^4$ *c/y* surviving imaging pictures with the [$AB\overline{C}$] coincidence are of course maintained. We consider realistic at this stage to bring the surviving sample to about 3 x $10^4$ *c/y*. With one fake event every 526 imaging pictures in the selected T300 volume, the resulting background is ≈ 76 *ev/y*. Any further possible reduction in the surviving numbers of imaging pictures will obviously correspondingly reduce the number of potential ν–e like background candidates.

2) We should record the positions and the fast timings $\tau_o(i)$ of all random muon tracks when crossing the walls of the 4π fast anti-coincidence [C] during the long LAr-TPC imaging. Each muon track may be determined correctly with the LAr-TPC imaging initiated with the specific drift time $\tau_o(i)$ of that signal and in agreement with the relevant rectangular fast detector counters. We remark that in order to generate an e.m. shower above the appropriate energy threshold (>200 MeV) the muon track must have sufficient energy (GeV), to traverse both in and out the Aluminium box. Showers are removed if the muon is at r ≤ 50 cm from the trajectory, assuming the drift time $\tau_o(i)$ of the muon track. The conversion probability is 0.98 and the subtracted volume ≈ 2.5 m³ or 3.5 ton for each removal[2].

---

[2] The removed areas are of different sizes and positions at each event.

Combining 1) and 2) the resulting background before inefficiencies is 76 x 0.02 =1.5 *ev/y*. Therefore with the help of these two additional conditions it should be possible to achieve the wanted goal of reducing realistically the very asymmetric e$^+$-e$^-$ pairs background to a final ν–e like background to less than a few *ev/y*.

The present ICARUS configuration is based on collecting the LAr light from each of the four T300 units. In view of the considerable longitudinal size of the T300 units (almost 18 m) it might be possible to further segment the light collection into N smaller, optically insulated units. This will reduce the number of potential muon tracks and of the γ-ray conversions in each imaging picture by the factor 0.19/N. As an example, for N = 4, a background count will be generated every 526 x 4 = 2100 imaging pictures. However the installation of these segmenting units is of considerable complexity and it may not be needed provided the above expectations 1) and 2) are not too far from truth.

The experimental determination of the surviving cosmic ray associated backgrounds is very complex and it requires also the collection of a substantial amount of additional calibration data with a pure cosmic ray trigger, namely outside the window of the 1.6 μs wide beam trigger.

As already apparent during the early experience of the LAr-TPC in Pavia during 2001, images at surface are generally dominated by a very considerable complexity. Even if the number of cosmic ray events can be substantially reduced with the help of the above methods and the introduction of the new anti-coincidence, events to be analysed are inevitably very many and require for each of them a large amount of information. Such a number of events cannot be scanned by hand and new automatic reconstruction methods must be developed, following the early experiences already initiated during the CNGS experiment.

*5. Conclusions.*

The LAr-TPC ICARUS detector has already operated successfully in the deep underground conditions of the Hall B of the GranSasso Laboratory at a distance of 730 km from the CERN-SPS. It has been a complete success, featuring a smooth operation, high live time, and high reliability. A total of about 3000 CNGS neutrino events have been collected and are being actively analyzed.

The idea of a short baseline experiment based on the LAr-TPC for a comprehensive search for sterile neutrinos was first presented by the ICARUS Collaboration at CERN in May 2009. During early 2011 the full proposal SPSC-P-345 was formulated for an experiment with the PS proton beam at 19.2 GeV/c and neutrino of an average momentum of 2 GeV/c, with two identical LAr-TPC at two different L/E$_ν$ distances of ≈100 m and 850 m. This proposal has evolved in 2012 in the joint ICARUS-NESSIE proposal SPSC-P-347 with the CERN-SPS and protons at 100 GeV/c at two different L/E$_ν$ distances of 350 m and 1800 m. The proposal was approved by the SPSC in 2013, but it has never been pursued, following the CERN decision that no accelerator neutrino beams should exist within Europe in the forthcoming future.

Therefore at the end of 2013 it has been proposed to locate the T600 detector at the Booster Neutrino Beam line (BNB) of FermiLab but with a much lower energy of 8 GeV. For an equivalent $L/E_\nu$ energy spectrum, the far detector will then be located at an approximate distance of about 600 m. As already mentioned, the experiment is presently under joint consideration at FNAL by three different teams based on as many as three distinct but similar LAr-TPC detectors at three different locations.

With a doubled mass, $2.2 \times 10^{20}$ p.o.t. (protons on target) and the 8 GeV FNAL Booster we may predict for a 100 % detection efficiency an intrinsic ν-e signal of 1780 $\nu_e/y$ and eventually an hypothetic LSND-like signal of about 495 $\nu_e/y$. The total number of beam associated neutrino ν–μ events visible in the LAr-TPC a the FNAL Booster will be about $5 \times 10^5$ *ev/y*, about ¼ of the one foreseen at the CERN-SPS and at about ¼ the neutrino average energy. As previously described, the actual detection efficiency after a very elaborate analysis should conservatively estimated between 30% and 40%, justifying the increases in the LAr mass.

As already pointed out and for the original ICARUS configuration, cosmic rays will make very relevant background contributions. Important additions must therefore be introduced. The main new element is a very efficient new 4π LAr anti-coincidence [C] covering all the peripheries of the walls of the detectors. The presence of this new component will affect the event rate with a number of reductions, with the expectation to bring the ν-e event rate to a value comparable with the expectation of the existence of a LSND signal. A summary of results is given in Table 1.

Such relatively simple 4π anti-coincidence [$AB\bar{C}$] is very effective in bringing the *direct* contribution due to cosmic rays under control. However, this reduction in the rate of background associated cosmic ray events will not be sufficient, since *any either beam related or otherwise recorded* "trigger" signal will automatically open a ms wide "imaging" window over which inevitably many additional and randomly distributed signals will be observed. Both Compton events and very asymmetric $e^+$-$e^-$ pairs may simulate an apparent singly ionizing e.m. track. While the Compton events may be effectively removed by the presence of a simulated "star" like recoil, the asymmetric $e^+$-$e^-$ pairs are much harder to remove.

The surviving ν–e background should be further reduced by at least two additional orders of magnitude, bringing it to the acceptable level with the help of the previously mentioned additional selection criteria 1) and 2).

Following 1), the probability of having in the same image more than one candidate event is small enough to be rejected. With a surviving sample of about $3 \times 10^4$ *c/y* "imaging" pictures, the remaining background is ≈ 76 *ev/y*.

Following 2), events are removed if a muon is at r ≤ 50 cm from the trajectory, reducing the ν–e background by a factor 50 but with but with a subtracted LAr volume of 3.5 ton for each visible muon.

*Table I. Summary of selections for ν–e events.*

| | | |
|---|---|---|
| Number of beam pulses | $5 \times 10^7$ | ev/y |
| Duration of each beam pulse | 1.6 | μs |
| Overall beam on time | 70 | sec/y |
| Distance from the target | 600 | m |
| Active mass LAr | 1000 | ton |
| νμCC event rate | $3.3 \times 10^5$ | ev/y |
| Quasi elastic events | 0.60 | |
| Single pion events | 0.30 | |
| Deep Inelastic (DIS) | 0.10 | |
| νμNC event rate | $1.1 \times 10^5$ | ev/y |
| Intrinsic ν-e beam rate | 1780 | ev/y |
| Possible LSND signal rate | 495 | ev/y |
| ν-e optimal energy window | 0.35-1.5 | GeV |
| Fraction νμCC in window | 0.81 | |
| Fraction ν-e in window | 0.72 | |
| Cosmic ray rate in 1 kton | 28.5 | kc/s |
| [AB] cosmic ray coincidence rate | $2.28 \times 10^6$ | ev/y |
| Dimensions of cells of 4π [C] anti-coincidence | 20 × 20 | cm$^2$ |
| Efficiency of 4π [C] anti-coincidence | 95 | % |
| Surviving [$AB\bar{C}$] cosmic ray rate | $1.71 \times 10^4$ | ev/y |
| γ-ray conversions in T300 for each imaging picture | 0.19 | |
| Imaging pictures in T300 for one Compton electron | 175 | i.p./ ev |
| The same, but with a visible recoil | >17'500 | i.p./ ev |
| Imaging pictures for 1 surviving pair + E < 5 MeV | 526 | i.p./ ev |
| Residuals of νμCC+νμNC + c.r. + .. and selection 1) | $3 \times 10^4$ | ev/y |
| Surviving fake events after selection 1) | 76 | ev/y |
| Fraction of e.m. events after selection 2) (r > 50 cm) | 0.02 | |
| C.R. related ν–e background after selections 1) + 2) | 1.5 | ev/y |
| Over-all detection related efficiencies | 0.4 | |

In conclusion and following Table 1 the cosmic ray related ν–e surviving backgrounds at a shallow depth, following the hardware improvements of the ICARUS detector and the selections 1) and 2) are presently estimated to be of the order of about 1.5 *ev/y*, with the expectations of combined detection inefficiencies and a remaining signal of about 40%.

Few events generated by neutrons, neutral kaons and so on are also present but they are most likely of much smaller concern, although they will have to be investigated.

Contributions due to single pion production are approximately one half of the quasi-elastic events, roughly divided in charged and neutral pion production. However, these events are much easier to recognize with a higher detection efficiency, because their higher complexities ($\mu^- p\pi^o$, $\mu^- n\pi^+$, $\mu^- p\pi^+$ and the corresponding $e^-$ channels) remove the bulk of the ν–e like background γ-ray induced effects due to the cosmic ray muons and their average energies are greater than the ones of the quasi elastic sample.

The contributions from the cosmic ray background require necessarily extended alternate running periods outside the beam rigger signal.

It will be virtually impossible to analyse visually such a large number of extremely complex events, following for each of them such elaborated procedures. A sophisticated automatic and very efficient method is therefore necessary and it has to be convincingly developed.

The modifications here described refer exclusively to the ICARUS programme. It is however likely that the introduction of similar new elements, and first of all the 4π proposed anti-coincidence counter should be necessary also for MicroBooNE, since located at only a slighter shorter distance (490 m vs. 600 m) from the target and perhaps also for the LAr1—ND programme.

Warm acknowledgements for their useful contributions and comments go to the whole ICARUS team and in particular to F. Pietropaolo, D. Gibin, S. Centro, A. Guglielmi, C. Montanari, and P. Sala.